\documentclass[a4paper,superscriptaddress,twocolumn,prb,showkeys,preprintnumbers,floatfix,showpacs]{revtex4-1}

\usepackage{epsfig}
\usepackage{amsmath}
\usepackage{amssymb}
\usepackage[colorlinks,bookmarks=false,citecolor=blue,linkcolor=red,urlcolor=blue]{hyperref}

\begin{document}

\title{Charge and Spin Fractionalization Beyond the Luttinger Liquid Paradigm}

\author{A. Moreno}
\affiliation{Institut f\"{u}r Theoretische Physik III, Universit\"{a}t Stuttgart, Pfaffenwaldring 57, D-70550 Stuttgart, Germany}

\author{A. Muramatsu}
\affiliation{Institut f\"{u}r Theoretische Physik III, Universit\"{a}t Stuttgart, Pfaffenwaldring 57, D-70550 Stuttgart, Germany}
\affiliation{Beijing Computational Science Research Center, Beijing, 100084, China }

\author{J. M.\ P.\ Carmelo}
\affiliation{Institut f\"{u}r Theoretische Physik III, Universit\"{a}t Stuttgart, Pfaffenwaldring 57, D-70550 Stuttgart, Germany}
\affiliation{Beijing Computational Science Research Center, Beijing, 100084, China }
\affiliation{Center of Physics, University of Minho, Campus Gualtar, P-4710-057 Braga, Portugal}

\date{\today}

\begin{abstract}
It is well established that at low energies 
one-dimensional (1D) fermionic systems are described by the Luttinger liquid (LL) theory, that predicts phenomena like spin-charge separation, and charge fractionalization into chiral modes. Here we show through the time evolution of an electron injected into a 1D $t$-$J$ model, obtained with time-dependent density matrix renormalization group, that a further fractionalization of both charge and spin takes place beyond the hydrodynamic limit. Its dynamics can be understood at the supersymmetric point ($J=2t$) in terms of the excitations of the Bethe-Ansatz solution. Furthermore we show that  fractionalization with similar characteristics extends to the whole region corresponding to a repulsive LL.
\end{abstract}

\maketitle

\section{Introduction}

There is a sustained interest in the physics of one-dimensional (1D) quantum systems due to recent experimental advances that allow to access exotic phenomena like spin-charge separation and charge fractionalization \cite{deshpande10}. At low energies these systems are well described by the Luttinger Liquid (LL) theory \cite{giamarchi04} that predicts two independent excitations carrying either only charge (holons) or only spin (spinons) and propagating with different velocities, and hence, spin-charge separation. Experimental evidences of its existence have been observed in quasi-1D organic conductors \cite{lorenz02}, semiconductor quantum wires \cite{auslaender05}, and quantum chains on semiconductor surfaces \cite{blumenstein11}. The LL theory also predicts the fractionalization of injected charge into two chiral modes (left- and right-going) \cite{pham00,trauzettel04,lebedev05,pugnetti09,das11}, a phenomenon recently confirmed experimentally \cite{steinberg08}. Along the experimental advances also 
theoretical progress was recently achieved 
pertaining extensions beyond the LL limit by incorporating nonlinearity of the dispersion, leading to 
qualitative changes in the spectral function 
\cite{imambekov09,imambekov09B,schmidt10,shashi11,carmelo05} and relaxation processes of 1D electronic systems \cite{barak10}. 

Here we show that fractionalization of charge and spin beyond the forms described by LL theory takes place when a spin-1/2 fermion is injected into a strongly correlated 1D system, namely the $t$-$J$ model. By studying the time evolution of the injected wavepacket at different wavevectors $k$, using time-dependent density matrix renormalization group (t-DMRG) \cite{white92, white93, white04,daley04,schollwoeck05,schollwoeck11} different regimes are obtained. When $k$ is close to the Fermi wavevector $k_F$, the known features from LL theory like spin-charge separation and fractionalization of charge into two chiral modes result. On increasing $k$, a further fractionalization of charge and spin appears, in forms that depend on the strength of the exchange interaction $J$ or the density $n$. Their dynamics can be understood at the supersymmetric (SUSY) point $J=2t$ in terms of charge and spin excitations of the Bethe-Ansatz solution \cite{bares90,bares91,bares92}. For the region of the phase diagram \cite{
ogata91,moreno11}, where the ground state corresponds to a repulsive LL, two qualitatively different regimes are identified: one regime with $v_s > v_c$ and another where $v_s < v_c$. Here $v_{c(s)}$ is the velocity of the excitations mainly carrying charge (spin). 
For $v_s > v_c$ and $k > k_F$ the spin excitation starts to carry a fraction of charge that increases with $k$ while $v_c$ corresponds to a wavepacket carrying only charge. For $v_s < v_c$ and $k > k_F$ the situation is reversed and the fastest charge excitation carries a fraction of spin that increases with $k$ while the wavepacket with $v_s$ carries almost no charge, i.e.\ in this case spin fractionalizes. 

The Hamiltonian of the 1D $t$-$J$ model is as follows,
\begin{eqnarray}
H & = & -t \sum _{i, \sigma}  
\left( \tilde{c}_{i,\sigma}^\dagger \tilde{c}^{}_{i+1,\sigma} + \mbox{h.c.} \right) 
\nonumber \\ & &
+ J \sum _{i} \left( {\vec S}_i \cdot {\vec S}_{i+1} - \frac{1}{4} n_i n_{i+1}   \right),
\label{Hamiltonian}
\end{eqnarray}
where the operator $\tilde{c}_{i,\sigma} ^\dagger$ ($\tilde{c}_{i,\sigma}$) creates (annihilates) a fermion with spin $\sigma= \uparrow$, $\downarrow$ on the site $i$. They are not canonical fermionic operators since they act on a restricted Hilbert space without double occupancy.  
${\vec S}_i = \tilde{c}^\dagger_{i,\alpha} \vec \sigma_{\alpha \beta} \tilde{c}^{}_{i,\beta}$ is the spin operator and $n_i= \tilde{c}_{i,\sigma}^\dagger \tilde{c}^{}_{i,\sigma} $ is the density operator.

We study the time evolution of a wavepacket, corresponding to a fermion with spin up injected into the ground state, by means of t-DMRG \cite{white92, white93, white04,daley04, schollwoeck05, schollwoeck11}. 
The state of a gaussian wavepacket 
$|\psi \rangle $ centered at $x_0$, with width $\Delta_x$ and average momentum $k_0$, is created by the operator $\psi^\dagger_\uparrow$ applied onto the ground state  $|G \rangle $:
\begin{equation}
 |\psi \rangle \equiv \psi^\dagger_\uparrow |G \rangle  = \sum_i \varphi _i \tilde{c}_{i \uparrow}^\dagger |G \rangle,
\label{e12}
\end{equation} 
with 
\begin{equation}
 \varphi _i = A e^{-(x_i-x_0)^2 /2\Delta_x} e^{ik_0 x_i}.
 \label{wavepaket2}
\end{equation} 
$A$ is fixed by normalization. The time evolved state
$|\psi(\tau ) \rangle $ by the Hamiltonian (\ref{Hamiltonian}) determines the spin ($s$) and charge ($c$) density relative to the ground state as a function of time $\tau$ measured in units of $1/t$ ($\hbar =1$),
\begin{equation}
\rho_\alpha(x_i,\tau) \equiv \langle \psi(\tau )|n_{i\alpha}|\psi(\tau )\rangle - \langle G|n_{i\alpha}|G\rangle,
\label{e9}
\end{equation}
where $\alpha=s,c$, $n_{i c} = n_{i\uparrow} + n_{i\downarrow}$, and $n_{i s} = n_{i\uparrow} - n_{i\downarrow}$.
Most of the numerical results were carried out on systems
with $L=160$ lattice sites, using 600 DMRG vectors (this translates into errors of the order of $10^{-4}$ in the spin and charge density up to times of $50/t$) and $\Delta_x=5$ lattice sites (which corresponds to a width $\Delta_k \sim 0.06\pi$ in momentum space). 

\section{Bethe-Ansatz solution}

At the supersymmetric (SUSY) point $J = 2t$ the 1D $t$-$J$ model can be solved exactly using Bethe-Ansatz \cite{bares91, bares92}. We consider here only the case of
zero magnetisation.
The solution is expressed in terms of two independent degrees of freedom, $c$ and $s$, related to two different kinds of pseudoparticles, with dispersion relations determined by 
\begin{eqnarray}
 \epsilon_ c (q) & =  & 4t\int_{-B}^{B}dr\,8r\,{[\bar{\Phi }_{s,c}\left(r, r_c(q)\right)-\bar{\Phi }_{s,c}\left(r, Q\right)]\over (1 + (2r)^2)^2},
\nonumber \\
\epsilon_ s (q) & = & -{4t \over{1+(2r_s(q))^2}}
\nonumber \\
& & + 4t\int_{-B}^{B}dr\,8r\,{[\bar{\Phi }_{s,s}\left(r, r_s(q)\right)-\bar{\Phi }_{s,c}\left(r, B\right)]\over (1 + (2r)^2)^2},
\label{dispersions}
\end{eqnarray}
where $q \in [-(\pi -k_{F}),(\pi -k_{F})]$, with $\alpha = c$ or $s$, 
$k_{F}=\pi n/2$, $n=N/L$, 
$N$ the number of electrons, and $L$ that of lattice sites. The range of momenta
for the excitations is later restricted to the occupied states for electron addition processes according to the pseudo-Fermi momenta given below, Eq.\ (\ref{PseudoFermiMomentum}).
The ground state rapidities $r_\alpha(q)$ (with $\alpha=c,s$) are defined in terms of their inverse functions
\begin{eqnarray}
q_c(r) & = & 4\int_{-B}^{B}dr'\,{\bar{\Phi }_{s,c}\left(r', r\right)
\over 1 + (2r')^2} \, , \hspace{0.25cm} r  \in [-\infty,\infty] \, ,
\nonumber \\
q_s(r) & = & 2\,\arctan(2r) 
\nonumber \\
& & + 4\int_{-B}^{B}dr'\,{\bar{\Phi }_{s,s}\left(r', r\right)
\over 1 + (2r')^2} \, , \hspace{0.25cm} r  \in [-\infty,\infty]  \, .
\label{GS-R-functions}
\end{eqnarray}

The functions $\bar{\Phi }_{\alpha,\alpha'}\left(r, r'\right)$ are the phase shifts defined by the following self-consistent integral equations
\begin{eqnarray}
\bar{\Phi }_{s,c}\left(r,r'\right) & = & -{1\over \pi}\arctan(2[r-r']) 
\nonumber \\
& & + \int_{-B}^{B}
dr''\,G(r,r'')\,{\bar{\Phi }}_{s,c}\left(r'',r'\right) \, ,
\label{Phisc-m}
\end{eqnarray}
and
\begin{eqnarray}
\bar{\Phi }_{s,s}\left(r,r'\right) & = & 
{1\over \pi}\arctan\Bigl(r-r'\Bigl) 
\nonumber \\
& & - {2\over{\pi^2}}\int_{-Q}^{Q} dr''{\arctan 
\Bigl(2[r''-r']\Bigr)\over{1+(2[r-r''])^2}} 
\nonumber \\
& & +\int_{-B}^{B} dr''\,G(r,r'')\,{\bar{\Phi }}_{s,s}\left(r'',r'\right) \, . 
\label{Phissn-m}
\end{eqnarray}
The kernel $G(r,r')$ reads,
\begin{eqnarray}
G(r,r') & = & - {1\over\pi}{1\over{1+(r-r')^2}}
\nonumber \\
& & + {4\over\pi^2}\int_{-Q}^{Q} dr'' {1\over{1+(2[r-r''])^2}}
{1\over{1+(2[r''-r'])^2}} 
\nonumber \\
& = & - {f(r,r')\over\pi}{1\over{1+(r-r')^2}},
\label{G}
\end{eqnarray}
where,
\begin{eqnarray}
f(r,r') & = & 1- {1\over 2}\left(t (r) + t(r') + {l (r) - l (r')\over 2(r-r')}\right) \, ,
\nonumber \\
t (r) & = & {1\over\pi}\sum_{j=\pm 1} j\,\arctan (2[r+j\,Q]) \, ,
\nonumber \\
l (r) & = & {1\over\pi}\sum_{j=\pm 1} j\,\ln (1+(2[r+j\,Q])^2) \, .
\label{t-l}
\end{eqnarray}

\begin{figure}[th!]
\begin{center}
$\begin{array}{cc}
\includegraphics[width=8.5cm]{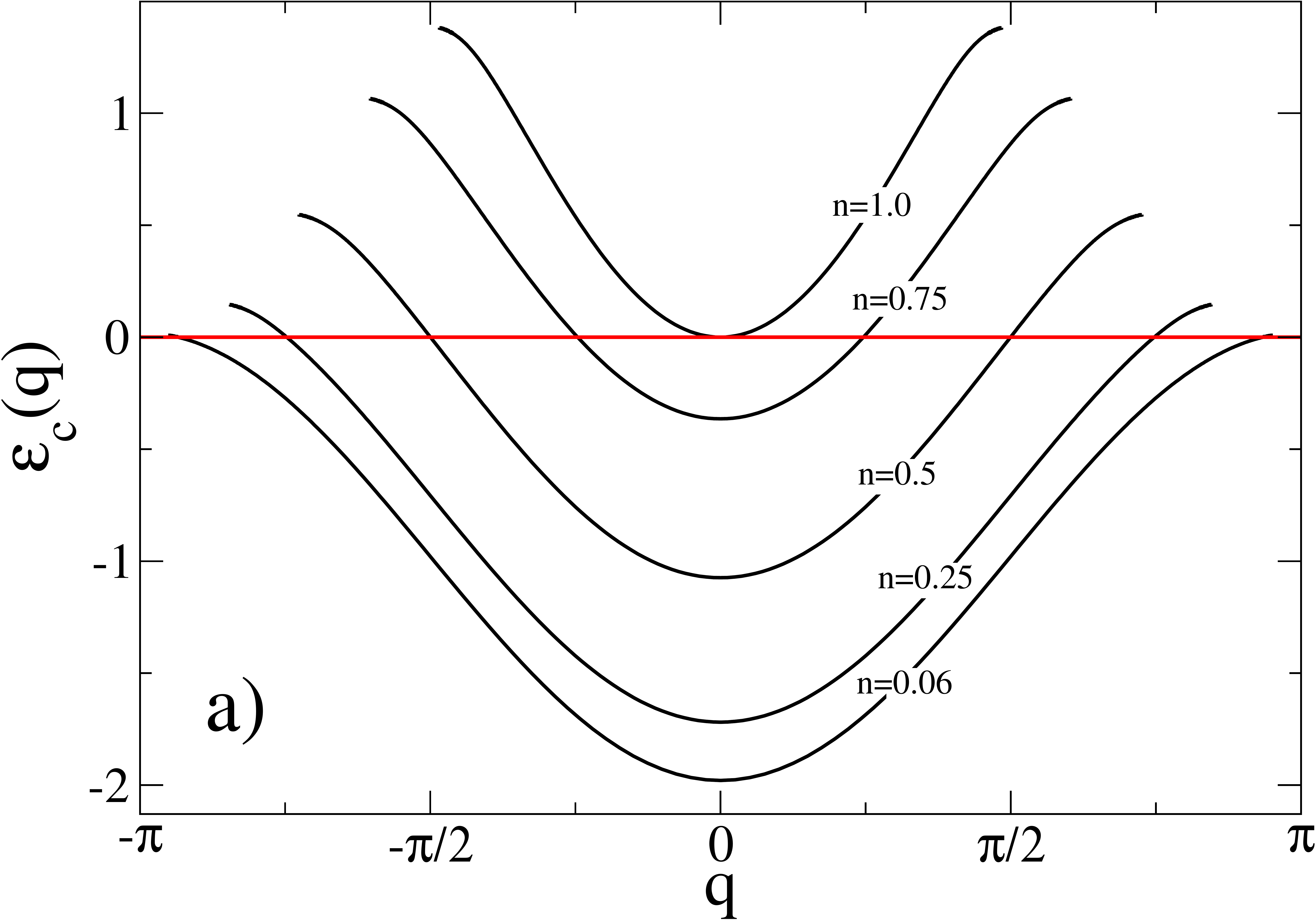} \\ 
\includegraphics[width=8.5cm]{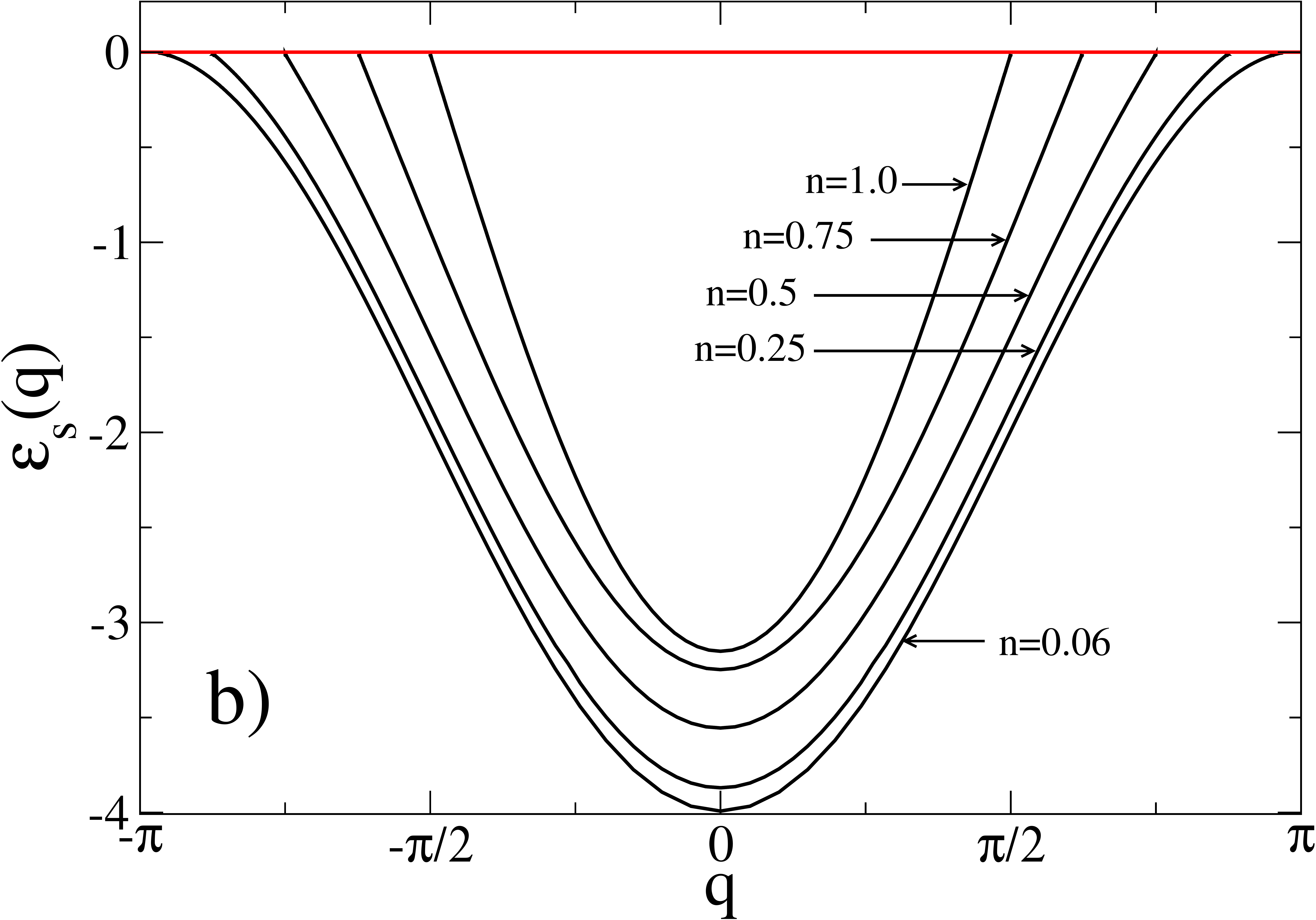}
\end{array}$
\end{center}
\caption{Solution of the dispersion relations (\ref{dispersions}) for different electronic density $n$. The ground state energy reference is defined such that $\epsilon_\alpha (\pm q_{F\alpha}) = 0$. a) $\alpha = c$ and b) $\alpha=s$.
The momentum range is given by $q \in [-(\pi -k_{F}),(\pi -k_{F})]$.}
\label{dispersions_cs}
\end{figure}
In the thermodynamic limit the ground state corresponds to symmetrical compact occupancies
of both $\alpha =c,s$ momentum bands (\ref{dispersions}) with Fermi momentum $q_{F\alpha}$ given by 
\begin{eqnarray}
\label{PseudoFermiMomentum}
q_{Fc} & = & (\pi -2k_F), \hspace{0.5cm} q_{Fs} =  (\pi -k_{F}) \, ,
\label{qcanGS}
\end{eqnarray}
respectively. The momenta of the states occupied in the ground state 
$q_c \in [-q_{Fc},q_{Fc}]$ and $q_s \in [-q_{Fs},q_{Fs}]$ refer to rapidity ranges $r\in [-Q,Q]$ and $r\in [-B,B]$, respectively, such that
\begin{equation}
r_c(\pm q_{Fc}) = \pm Q \, ; \hspace{0.5cm} r_s(\pm q_{Fs}) = \pm B, 
\label{RqcanGS}
\end{equation}
where $Q$ and $B$ are 
obtained by solving self-consistently the normalization conditions given by the following integral equations 
\begin{eqnarray}
\pi -2k_F & = & 4\int_{-B}^{B}dr\,{\bar{\Phi }_{s,c}\left(r, Q\right)
\over 1 + (2r)^2} \, , 
\nonumber \\
\pi -k_{F} & = & 2\,\arctan(2B) + 
4\int_{-B}^{B}dr\,{\bar{\Phi }_{s,s}\left(r, B\right)
\over 1 + (2r)^2} \, .   
\label{QB-GS-R-functions}
\end{eqnarray}

We proceed by solving Eqs.\ (\ref{Phisc-m}) and (\ref{Phissn-m}) assuming that $Q$ and $B$ are known and then we use Eqs.\ (\ref{QB-GS-R-functions}) and $k_{F}=\pi n/2$ to find
the corresponding electronic density $n$. In Fig.\ \ref{dispersions_cs} we show the resulting dispersion relations for different values of $n$.
The ground state energy reference is defined such that $\epsilon_\alpha (\pm q_{F\alpha}) = 0$. 
The dispersions plotted in Fig.\ \ref{dispersions_cs}
are the ones entering the calculation of velocities discussed in the next section.

\section{Simulations at the SUSY point}
We discuss first the time evolution of a wavepacket at the SUSY point $J = 2t$, since here we will be able to identify the different portions in which the wavepacket splits on the basis of the Bethe-Ansatz solution. 
\begin{figure}[th!]\relax 
\centerline{\includegraphics[width=8.6cm]{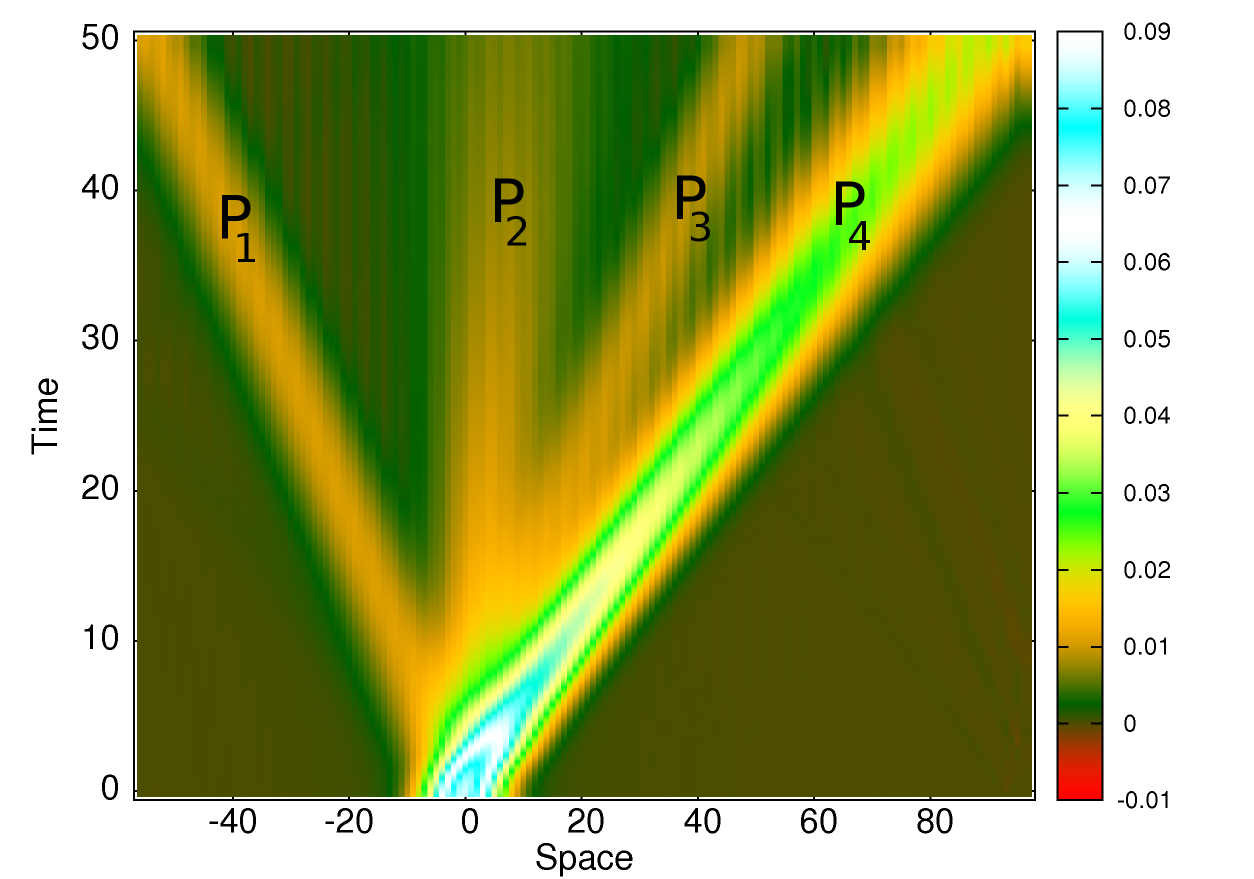}}
\caption{(color online). Time evolution of $\rho_c (x_i,\tau)$ for a wavepacket initially at $x=0$, with momentum $k=0.7\pi$, at density $n=0.6$, and $J=2t$.
Charge fractionalizes into four wavepackets, one to the left($P_1$) and the
rest($P_2$, $P_3$, $P_4$) to the right. $P_1$ and $P_3$ have the same charge and speed but opposite velocities.}
\label{J2TimeSpaceCharge}
\end{figure}
Figure \ref{J2TimeSpaceCharge} shows the time evolution of $\rho_c (x_i,\tau)$ for a density of $n=0.6$.
The momentum of the injected fermion is $k=0.7\pi$, i.e.\ midway between $k_F = 0.3 \pi$ and the zone boundary. The charge (i.e.\ $\rho_c$) splits into four fractions, one portion traveling to the left and the rest doing so to the right. A splitting into chiral modes is expected in the frame of LL theory \cite{pham00}, where for an injected right-going fermion, a splitting $Q_\alpha^{(\pm)} = (1\pm K_\alpha)/2$ (where $K_\alpha$ is the so-called LL parameter and '+' ('-') corresponds to the right (left) propagating part) is predicted. The amount of charge (i.e.\ the integral of the wavepacket over its extension) corresponding to the portion denoted $P_1$ is  $Q_{c}^{(-)} \sim 0.1$. This value is independent of the momentum of the injected fermion, and 
agrees well with the prediction of LL theory, since for the parameters in this case, $K_c \sim 0.8$ \cite{moreno11}. 
However, at long enough times, a further splitting of the right-going charge is observed (wavepackets $P_i$ with $i=2,3,4$), beyond the prediction of the LL theory. 

\begin{figure}[th!]\relax 
\centerline{\includegraphics[width=8.6cm]{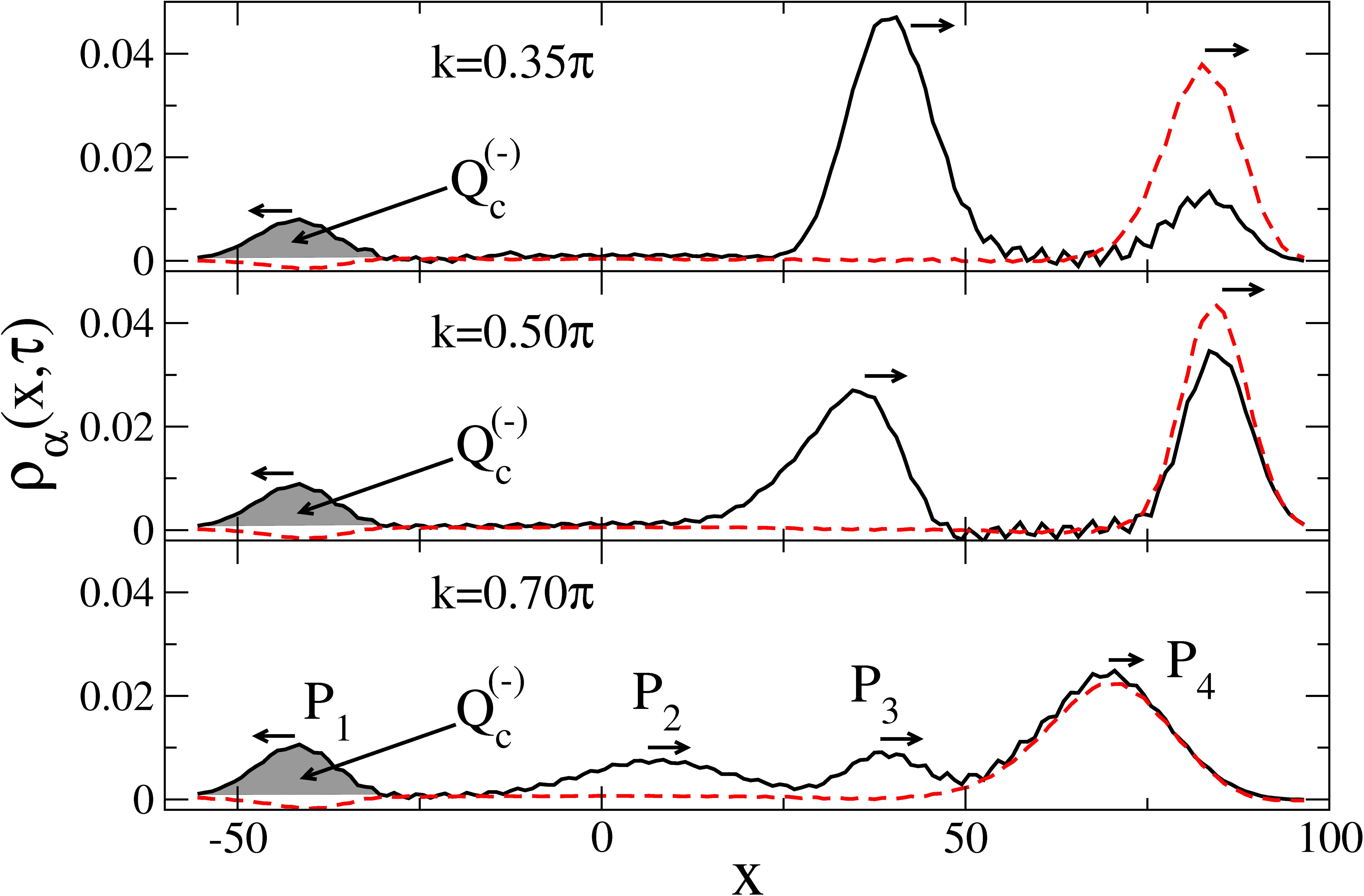}}
\caption{(color online). Charge ($\rho_c (x_i,\tau)$, full line) and spin ($\rho_s (x_i,\tau)$, dashed line) densities
for $J=2t$, $n=0.6$, at time $\tau=40$, for different values of the momentum of the injected fermion. $Q^{(-)}_c$ denotes the charge of the left going wavepacket. Its value ($\sim 0.1$) remains unchanged in all three panels.}
\label{J2Qs}
\end{figure}
\begin{figure}[th!]
\begin{center}
\includegraphics[width=8.5cm]{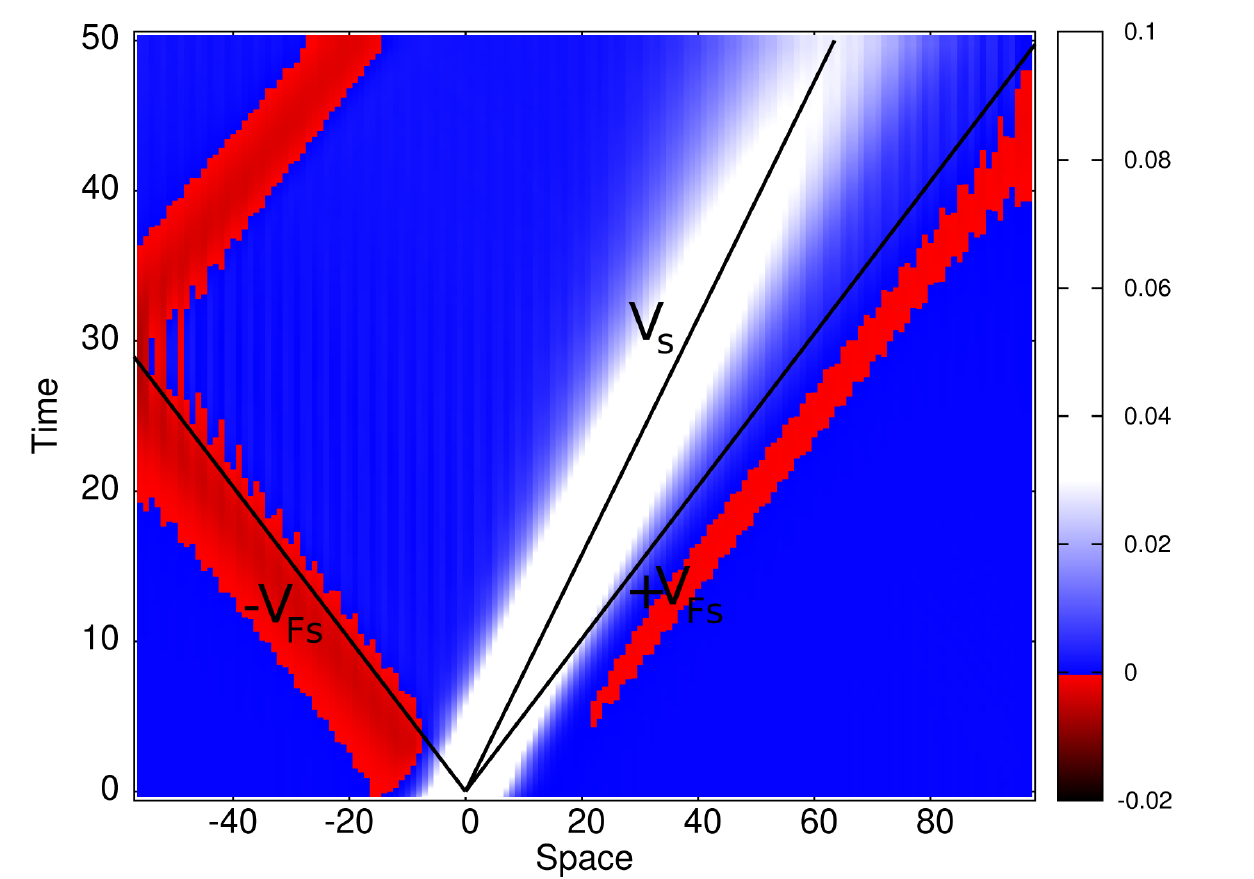} 
\end{center}
\caption{Time evolution of $\rho_s (x_i,\tau)$ for a wavepacket with the same parameters as Fig.\ \ref{J2TimeSpaceCharge}. The small depressions in spin density (red) have a slope in a time-space diagram corresponding to the Fermi velocity $v_{Fs}$ of $\epsilon_s$. $v_s$ denotes the velocity of the wavepacket $P_4$ in Fig.\ \ref{J2Qs}, that contains most of the spin density.}
\label{KsQuestion}
\end{figure}

Figure \ref{J2Qs} displays both $\rho_c (x_i,\tau)$ 
(full line) and $\rho_s (x_i,\tau)$ (dashed line) for different values of the initial momentum of the injected wavepacket. The arrows indicate the direction of motion of each packet.
As opposed to $\rho_c$, $\rho_s$ does not split.
In a SU(2) invariant LL $K_s = 1$ \cite{giamarchi04}, and, assuming that the left going wavepacket for spin is described by LL theory as in the case of charge, we would have
$Q_s^{(-)} = 0$, i.e.\ no left propagating part 
is expected for the spin density. (However, a small depletion in $\rho_s$ appears traveling to the left, which would correspond to $K_s \gtrsim 1$. Similar findings were presented recently \cite{soeffing12} and attributed to finite-size effects that require exponentially large systems in order to recover $K_s=1$. Figure \ref{KsQuestion} displays the time evolution of the spin densities at the SUSY point, in order to explicitly show that this small depression in spin-density moves with the Fermi velocity $v_{Fs} \simeq 2ta$, where $a$ is the lattice constant set to one, at the pseudo-Fermi sea in Fig.\ \ref{dispersions_cs} (b)).  
Moreover, part of the charge ($P_4$) is accompanying the spin, such that spin-charge separation does not appear to be complete. The amount of charge accompanying the spin increases as the momentum of the injected fermion approaches the zone boundary. These results make already evident that injecting a fermion at a finite distance from the Fermi energy leads to fractionalization of charge beyond the expectations from the LL theory. 

In order to understand the new forms of fractionalization that go beyond the LL frame, we consider 
the excitations corresponding to one-particle addition processes, whose energies can be obtained 
from the Bethe-Ansatz solution \cite{carmelo05,carmelo06b}.  
When adding an electron with momentum $k$, the single particle excitation energy is given by $\omega (k) = -\epsilon_c (q_c) -\epsilon_s (q_s)$, where $\epsilon_c(q_c)$ and $\epsilon_s(q_s)$ are the dispersion relations (\ref{dispersions}) of the excitations for charge and spin, respectively, and the momenta are related to the momentum of the incoming particle as follows: $k =  \pm 2k_F - q_c - q_s$, where $q_c \in [-q_{Fc},q_{Fc}]$, and $q_s \in [-q_{Fs},q_{Fs}]$, with $q_{Fc}$ and $q_{Fs}$ the pseudo-Fermi momenta for the excitations for charge and spin, given in Eqs.\ (\ref{qcanGS}), respectively \cite{bares92}.  

\begin{figure}[th!]\relax 
\centerline{\includegraphics[width=8.6cm]{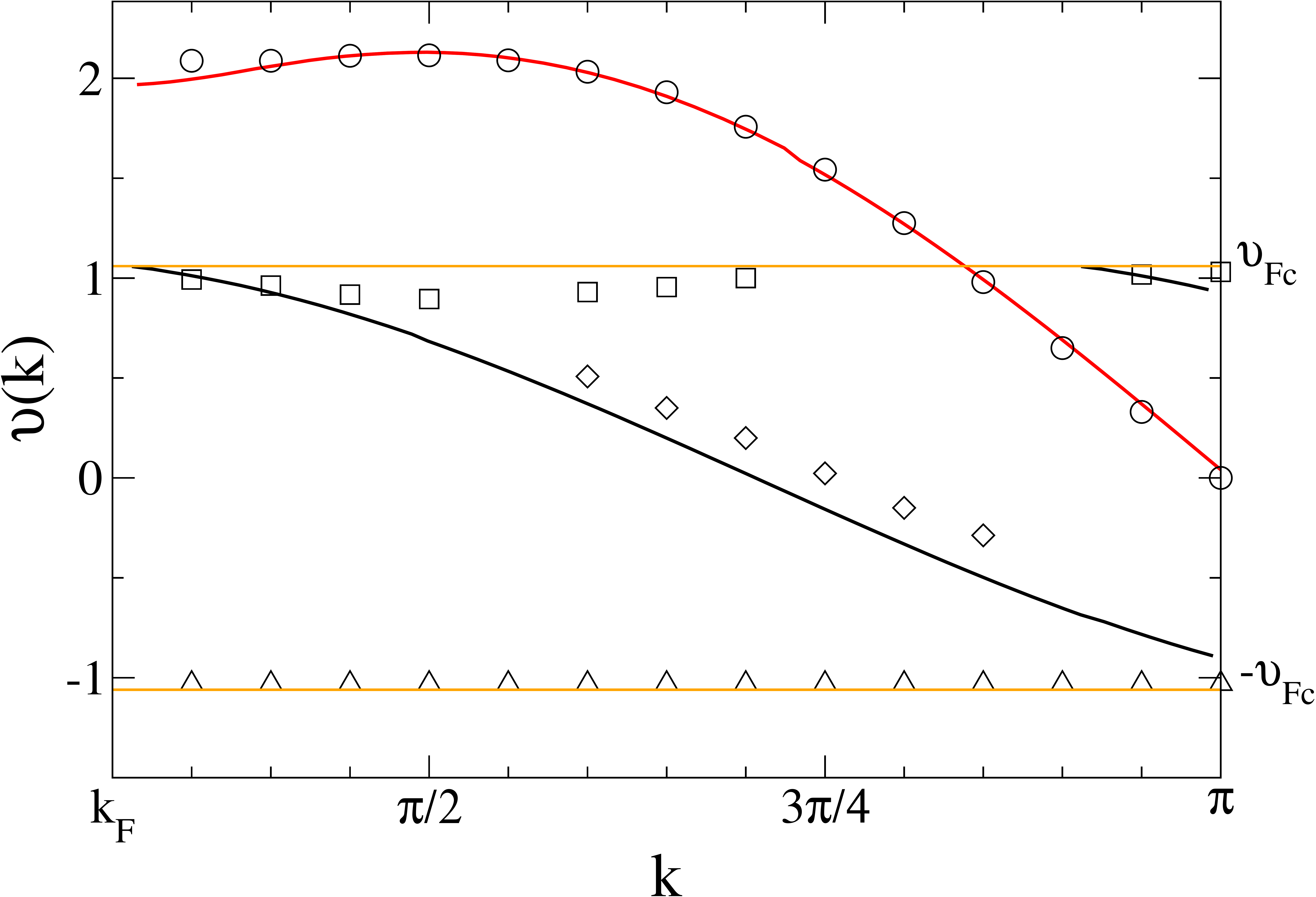}}
\caption{(color online). Full lines: derivatives $v_\alpha (k) = \partial \epsilon_\alpha (k) /\partial k$  
of the dispersions obtained 
by the Bethe-Ansatz solution. The symbols correspond to   
the velocities of the different wavepackets identified in Fig.\ \ref{J2Qs}: triangles($P_1$), diamonds($P_2$), squares($P_3$) and circles($P_4$). The orange horizontal lines stand for the Fermi velocity $v_{Fc} = \partial \epsilon _c (q_{Fc}) /\partial q$ given by the Bethe-Ansatz solution.}
\label{fig3}
\end{figure}
Figure \ref{fig3} displays the velocities obtained from t-DMRG for the different wavepackets (symbols) compared to those obtained from Bethe-Ansatz (full lines), as a function of the momentum of the injected fermion.
The velocity of each $P_i$ is extracted by measuring the position of the maximum of the packet at the most convenient time, i.e.\ at that time where we can resolve $P_i$ and 
the spreading of one packet does not destroy the other packets.
The wavepackets $P_1$ (triangles) and $P_3$ (squares) have opposite directions, but the same speed and charges $Q_{c}^{(-)} \simeq Q_{c3}^{(+)}$, where the charges for the $(+)$ branch 
are labeled by  
an index corresponding to the respective wavepackets. 
The velocity of the wavepacket $P_4$ (circles) agrees almost perfectly with the one corresponding to spin excitations. Its determination is best since it is the fastest wavepacket, such that it can be easily discerned from the rest. The velocity of the remaining wavepacket, $P_2$ (diamonds), is more difficult to assess, since 
it overlaps at the beginning with other ones. Nevertheless, its velocity closely follows the one of charge excitations. The wavepackets just described deliver a direct visualization of the excitations appearing in the Bethe-Ansatz solution, where only two different kinds of particles are involved: the $c$ and $s$ pseudoparticles with their associated bands. 
The excitation associated with spin involves one hole in the $c$ band with fixed momentum $q_{Fc}$ and one hole in the $s$ band with momentum $q_s$, where
$q_s = \pm 2k_F -q_{Fc} -k$ \cite{bares92}.
In fact, the velocities of $P_1$ and $P_3$ correspond to the group velocity at both pseudo-Fermi momenta  $\pm q_{Fc}$,
indicating that these wavepackets  
correspond to low energy excitations. This explains the fact that $Q^{(-)}_c$ is well described by LL theory in spite of the fermion being injected at high energy, and supports the assumption that the same applies to a left going wavepacket for spin (see Fig.\ \ref{KsQuestion}). 
Furthermore, as shown in Fig.\ \ref{fig3}, the velocity of those fractions is independent of the momentum of the injected fermion, in agreement with the picture given by Bethe-Ansatz. The dispersion of the hole in the $s$-band gives rise to the velocity displayed by the red line in  Fig.\ \ref{fig3}. 
Similarly, the $c$ line (black line in Fig.\ \ref{fig3}) involves one hole in the $s$ band with fixed momentum $q_{Fs}$ and one hole in the $c$ band with momentum $q_c$ determined in terms of $k$ by $q_c = \pm 2k_F -q_{Fs} -k$. Using the same argument as for
the $s$ line we can associate the $P_2$ packet (diamonds) with the $c$ pseudoparticle. However, in this case we cannot observe wavepackets associated with spin and 
velocities corresponding to the group velocity at the pseudo-Fermi momenta $\pm q_{Fs}$. 
We understand this as due to the fact that $K_s=1$, by analogy to what we observe in the 
$K_c=1$ case. On the SUSY point this case is reached in the limit of vanishing density, where the system can be described by a Fermi gas. Hence, fractionalization is absent in this limit.
\begin{figure}[th!]
\begin{center}
\includegraphics[width=8.5cm]{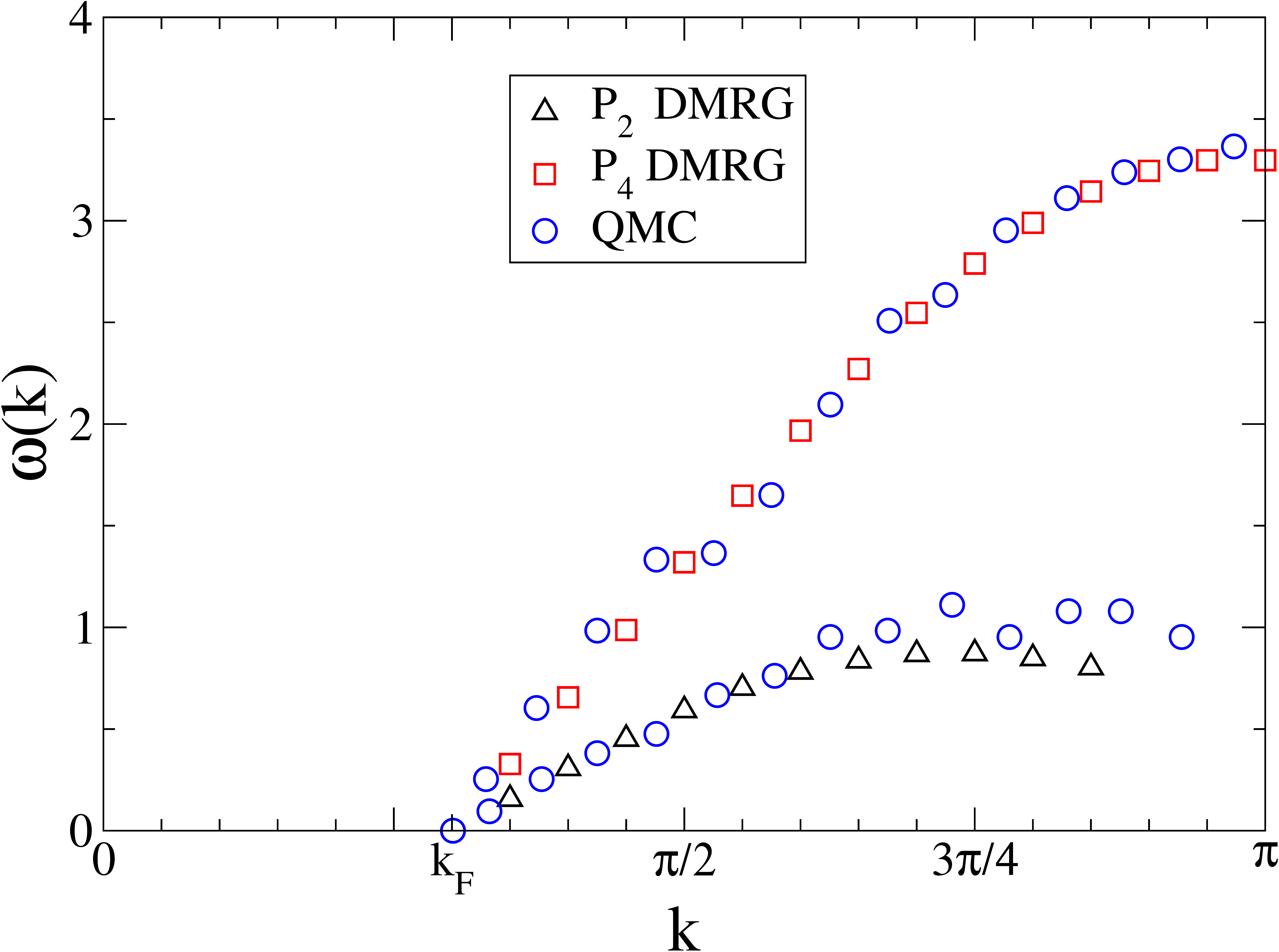} 
\end{center}
\caption{Locus of the highest weight features of the electron addition part of the spectral function of the $t$-$J$ model from quantum Monte Carlo results \cite{lavalle03} and the energies obtained by integrating the velocities of the wavepackets $P_2$ and $P_4$.}
\label{QMC-DMRG}
\end{figure}
In Fig.\ \ref{QMC-DMRG} we show finally, a comparison of dispersions with highest weight in the spectral function obtained from quantum Monte Carlo (QMC) simulations for the one-dimensional $t$-$J$ model \cite{lavalle03} and the energies obtained from t-DMRG by integrating the velocities between $k_F$ and $k$ with the zero of energy at $k_F$. While the dispersions obtained in QMC simulations can be well reproduced by the velocities obtained from the wavepackets $P_2$ and $P_4$ from t-DMRG, 
given the discretization errors in integrating the velocities, and uncertainties from the analytic continuation in QMC, no direct access to the wavepackets $P_1$ and $P_3$ is possible from the spectral function. Their contribution to the spectral function is contained in the intensities of the spectrum, but no distinct feature allows to extract them from it.

\section{Away from the SUSY point}

Next we depart from the SUSY point and examine how fractionalization takes place in the region of the phase diagram where the ground state corresponds to a LL with $K_c < 1$.
\begin{figure}[th!]\relax 
\centerline{\includegraphics[width=8.6cm]{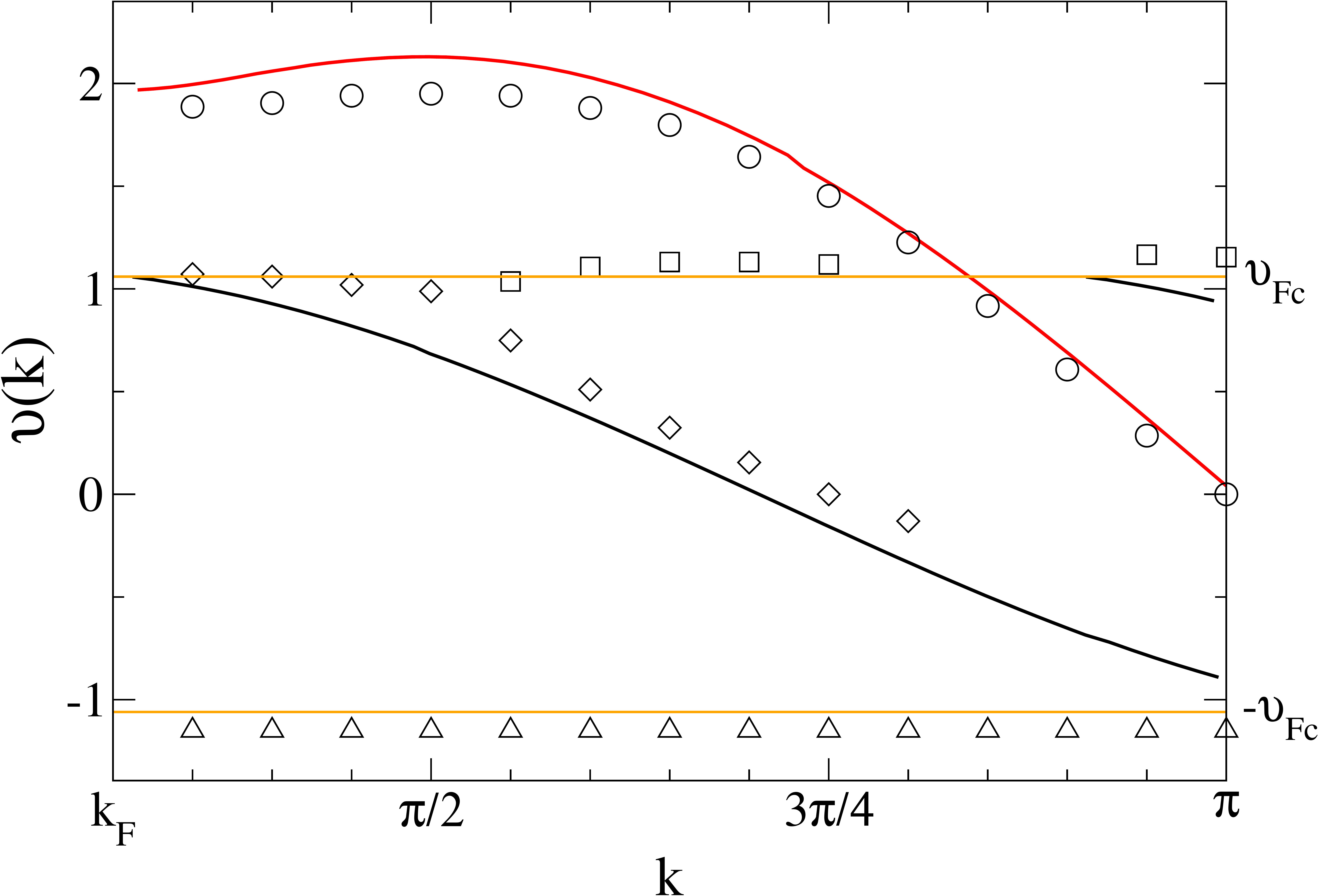}}
\caption{(color online). As in Fig.\ \ref{fig3} but for $J=1.75t$. The fullines correspond to the SUSY Bethe ansatz.}
\label{J175n06}
\end{figure}
Figure \ref{J175n06} shows the velocity of the different fractions at $J=1.75 t$, where a slight decrease (increase) in the velocity of the spin (charge) fraction can be observed. As shown in Fig.\ \ref{awayJ2}, essentially the same features are observed as at the SUSY point both for $J > 2t$ and $J< 2t$. 
\begin{figure}[th!]\relax 
\centerline{\includegraphics[width=8.6cm]{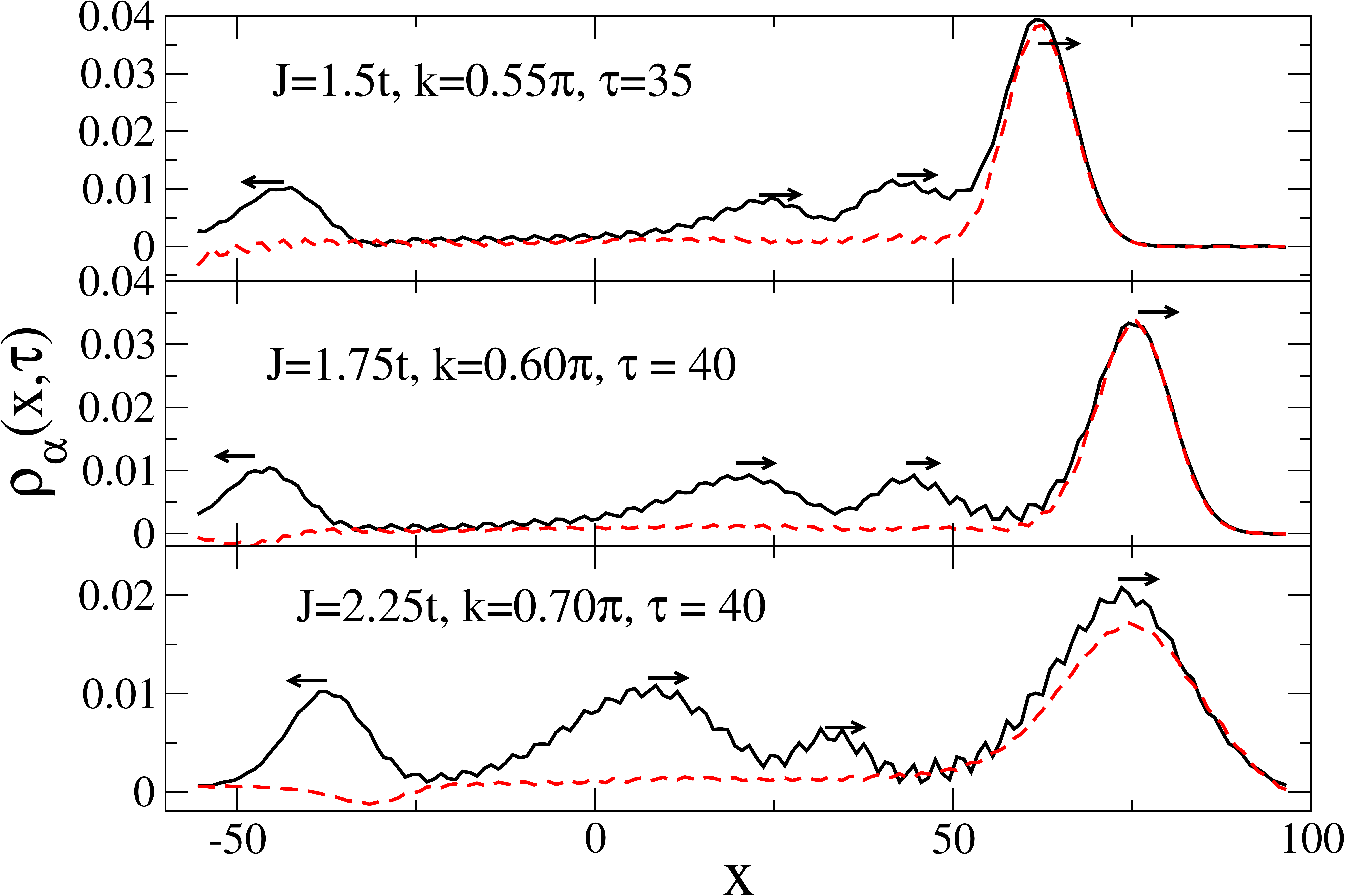}}
\caption{(color online). Fractionalized wavepackets for different values of $J/t$ away from the SUSY point, at a density
$n=0.6$. As in the SUSY case, charge fractionalizes into four pieces, while spin does not, and carries an appreciable amount of charge.
}
\label{awayJ2}
\end{figure}
In all the cases shown in Fig.\ \ref{awayJ2}, where the velocity of spin excitations ($v_s$) remains higher than that
of charge excitations ($v_c$) in most parts of the Brillouin zone,
spin does not fractionalize, as opposed to charge, so that the interpretation derived from Bethe-Ansatz remains valid over an extended region of the phase diagram: charge splits into four portions of which one travels with the spin wavepacket, and two have the same speed but opposite group velocity which
does not depend on the momentum of the injected fermion.  
It is tempting to assign those excitations to states at a pseudo-Fermi surface for charge excitations. For smaller values of $J/t$ than those in Fig.\ \ref{awayJ2}, $v_s$ becomes smaller than $v_c$. Figure \ref{SpinFrac} shows that for $v_s < v_c$ the role of spin and charge wavepackets experience a change with respect to fractionalization. 
\begin{figure}[th!]\relax 
\centerline{\includegraphics[width=8.6cm]{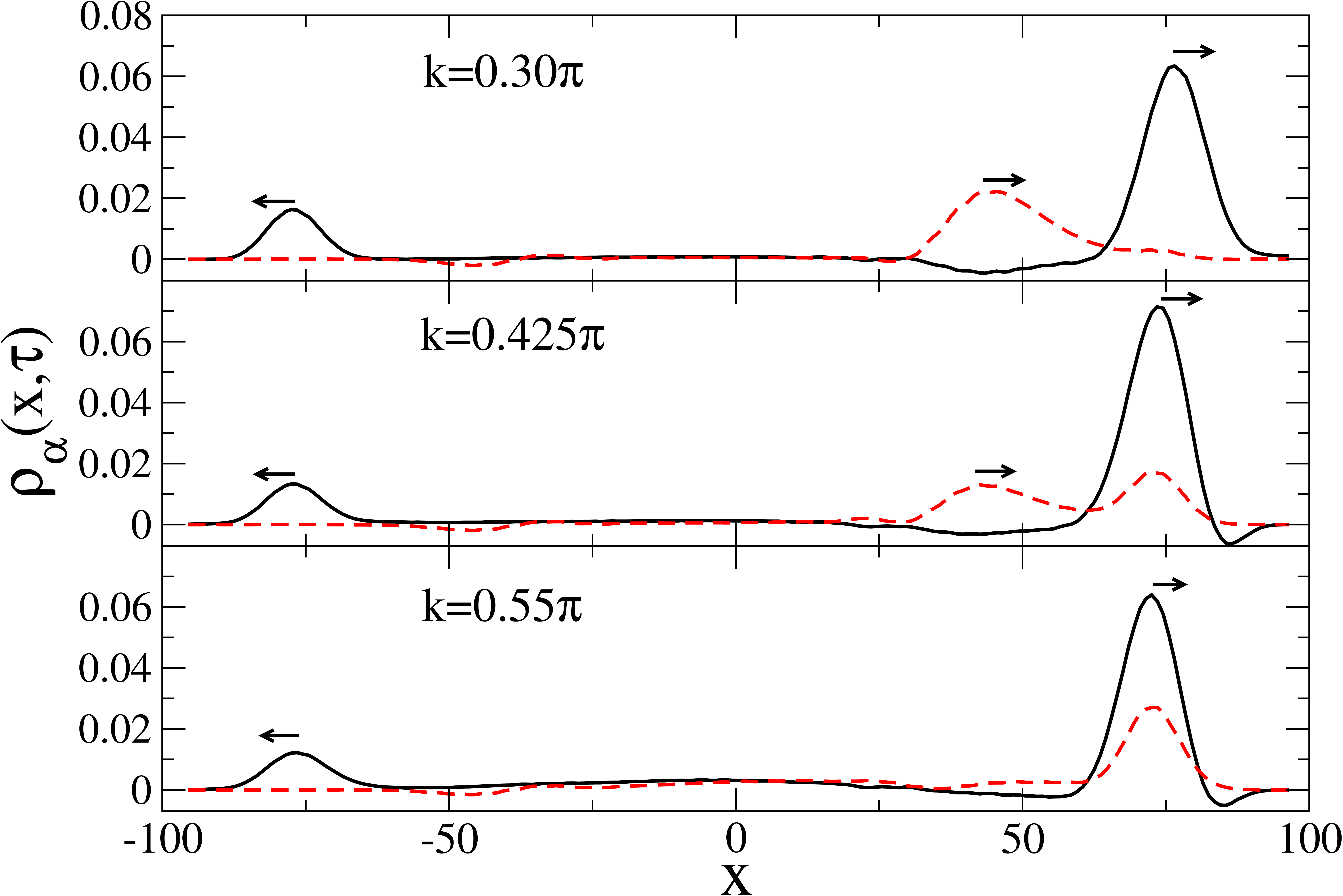}}
\caption{(color online). Fractionalized wavepackets at $J/t=1$, $n=0.5$, $\tau=50$ and $L=200$. In this case, where $v_s < v_c$, fractionalization of the spin density is observed.}
\label{SpinFrac}
\end{figure}
In this case it is the spin density that splits into two fractions, one attached to the fastest fraction of charge, an another one left behind. Again, this new feature is not predicted by LL theory.
However,  
no left propagating fraction of spin could be observed (excepting the small depression due to finite-size effects). We therefore expect 
that this is due to SU(2) symmetry and the fact that in this case $K_s=1$. As shown in Fig.\ \ref{SpinFrac}, the amount of spin accompanying the charge increases as the momentum of the injected fermion increases.
While such a phenomenon may suggest as in the lowest panel of Fig.\ \ref{awayJ2} a total recombination of charge and spin as the energy increases, it is not total, since still a fraction of charge goes to the left, without accompanying spin. 
\begin{figure}[th!]
\begin{center}
\includegraphics[width=8.5cm]{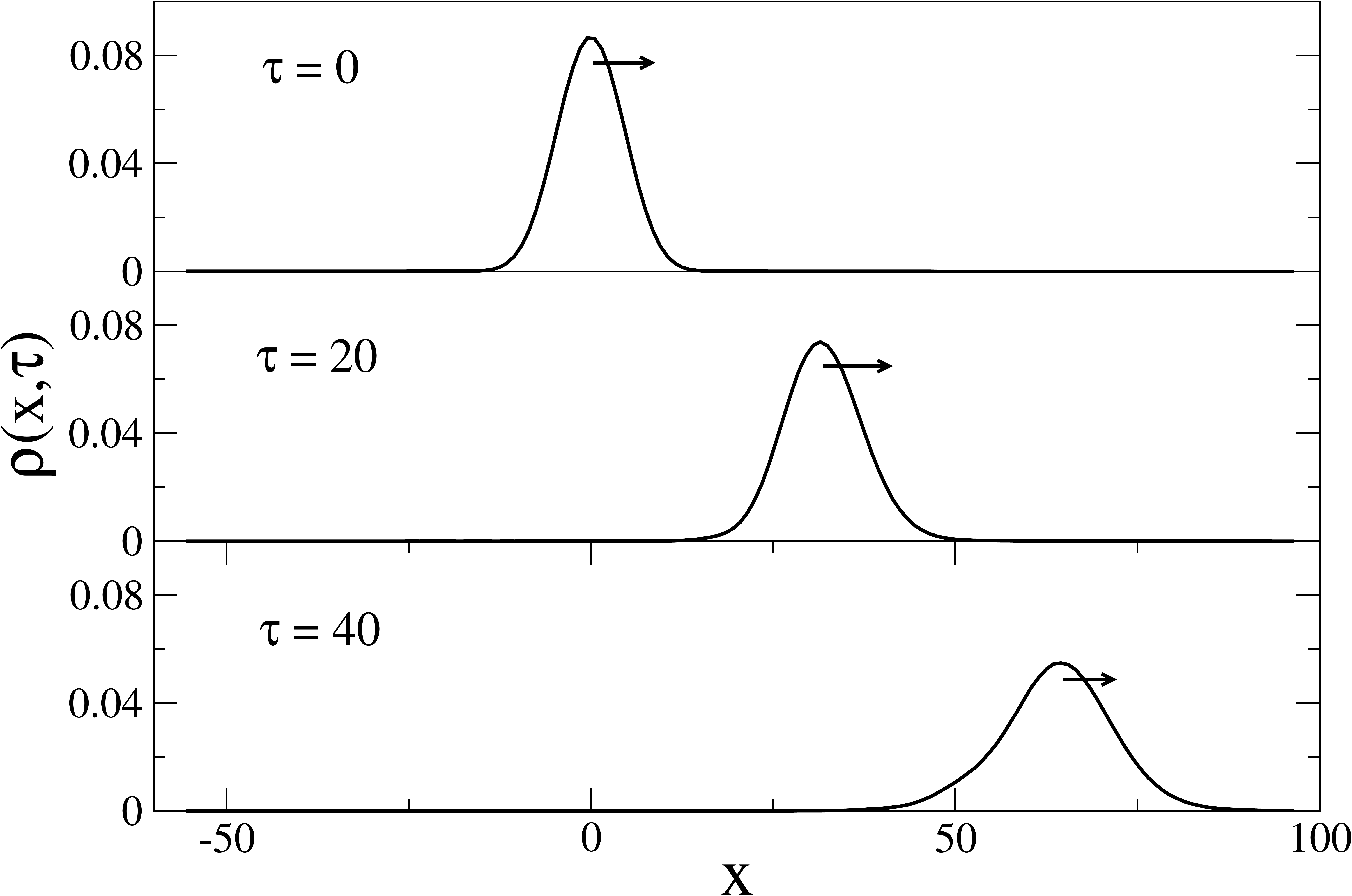} 
\end{center}
\caption{Free expansion of noninteracting spinless fermions at different times, where the wavepacket disperses for increasing time. The momentum of the injected fermion is $k = 0.7 \pi$ and $n=0.5$.}
\label{FreeCase}
\end{figure}

It is important to check whether the observed fractionalizations in fact correspond to elementary excitations and not to other effects like the band curvature or the forbidden double occupancy. In Fig.\ \ref{FreeCase} we present the result for a non-interacting system, where, as expected no fractionalization takes place. The effect of the band curvature is merely to give a dispersion of the wavepacket, as taught in elementary quantum mechanics for a free particle.
The $t$-$J$ model reduces in the limit $J \rightarrow 0$ to the Hubbard model for $U \rightarrow \infty$, where the ground-state wavefunction can be factorized in a part related to charge and another related to spin \cite{ogata90}, such that spin-charge separation can be expected at all energies.

\begin{figure}[th!]
\begin{center}
\includegraphics[width=8.5cm]{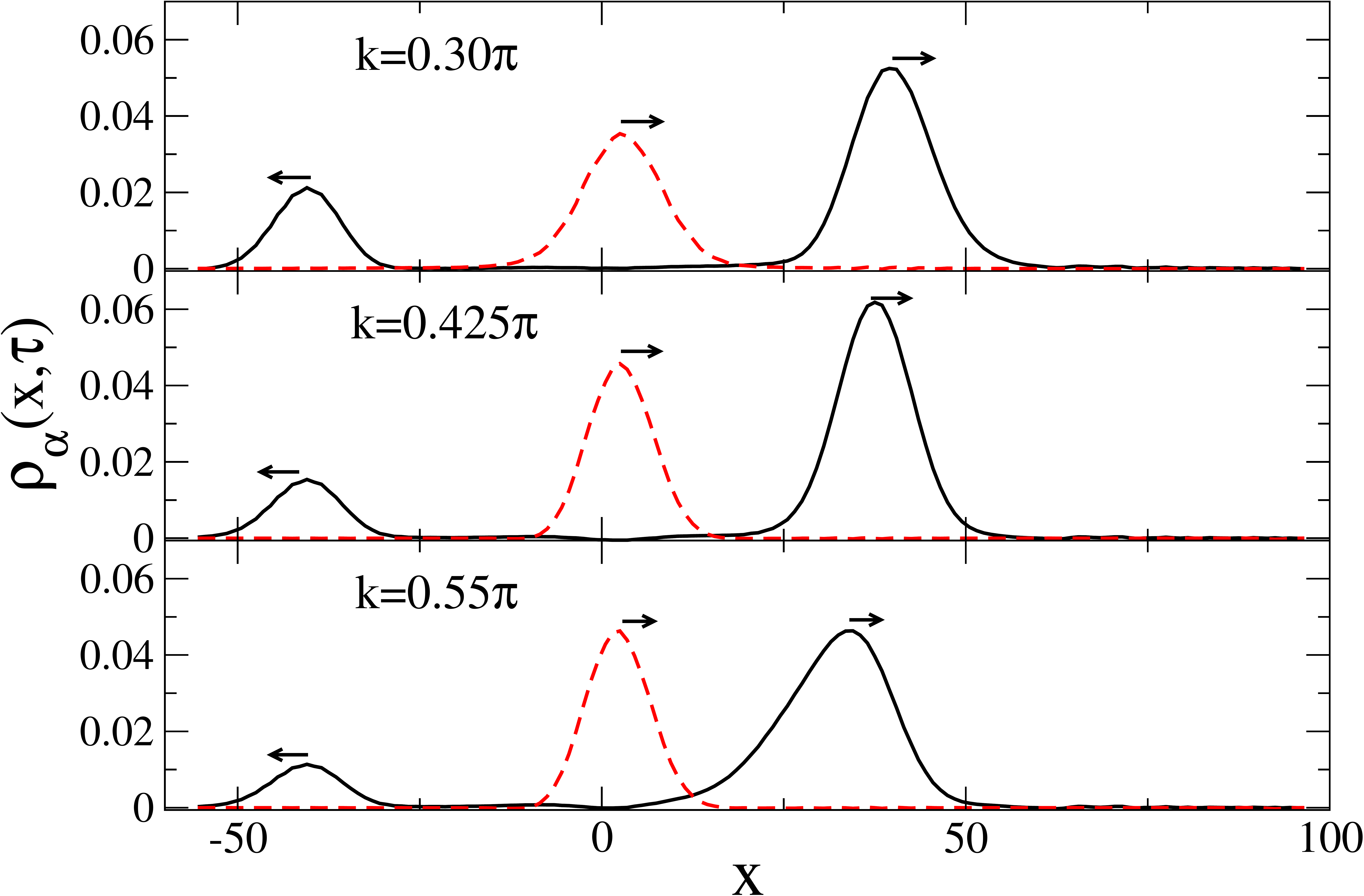} 
\end{center}
\caption{Fractionalized wavepackets at $J/t=0.1$, $n=0.5$, $\tau=50$ and $L=200$ to be compared with Fig.\ \ref{SpinFrac}. As expected for the limit $U \rightarrow \infty$ of the Hubbard model, spin-charge separation is observed for the different wavevectors of the injected fermion.}
\label{JToZero}
\end{figure}

Figure \ref{JToZero} shows the wavepackets evolving at $J=0.1 t$ for the same parameters as in Fig.\ \ref{SpinFrac}. Here, spin-charge separation can be observed for the different wavevectors of the injected fermion. Hence, the fractionalizations and recombinations observed are not just a consequence of forbidden double occupancy or band curvature.

\section{Conclusions}

In summary, we have shown through the time evolution of an injected spinfull fermion onto the $t$-$J$ model, that charge and spin fractionalization occurs beyond the predictions of the Luttinger liquid theory. A comparison with results from Bethe-Ansatz allowed to identify charge and spin excitations that split into components at high and low energies. The components at high energy reveal the dispersion $\epsilon_c$ and $\epsilon_s$ of charge and spin excitations, respectively. The components at low energy have a velocity that does not depend on the momentum of the injected fermion and are very well described by states at the pseudo-Fermi momenta of the charge excitation. This picture can be extended to a wide region in the phase diagram of the $t$-$J$ model as long as the ground state corresponds to $K_c < 1$ and $v_s > v_c$. In this region fractionalization is observed only in the charge channel. However, for $v_c > v_s$, a region that develops for $J/t$ below  $\sim 1.5$, the spin density shows 
fractionalization. All over, the fastest excitation is accompanied by the complementary one, such that spin-charge separation is for them only partial. The other fractions present an almost complete spin-charge separation. 

Finally, we would like to remark, that the time evolution 
leads to a direct visualization of all fractions stemming from an injected fermion in contrast to
the one-particle spectral function, where only the fractions $P_2$ and $P_4$ can be identified \cite{lavalle03}, 
but not those propagating at the pseudo-Fermi points.

A.\ M.\ and A.\ M. acknowledge support by the DFG through SFB/TRR 21.
A. M. and J. M. P. C. thank the hospitality and support of the Beijing
Computational
Science Research Center, where part of the work was done. J. M. P. C. thanks
the
hospitality of the Institut f\"ur Theoretische Physik III, Universit\"at
Stuttgart, and
the financial support by the FEDER through the COMPETE Program, Portuguese
FCT
both in the framework of the Strategic Project PEST-C/FIS/UI607/2011 and
under SFRH/BSAB/1177/2011, German transregional collaborative research
center
SFB/TRR21, and Max Planck Institute for Solid State Research.  A.M.\ thanks the KITP for
hospitality. This research was
supported in part by the National Science Foundation under Grant No.\ NSF
PHY11-25915.


\begin{thebibliography}{1}

\bibitem{deshpande10}
V.~V. Deshpande, M. Bockrath, L. Glazman, and A. Yacobi, Nature {\bf 464},  209
   (2010).

\bibitem{giamarchi04}
T. Giamarchi, {\em Quantum Physics in One Dimension} (Clarendon Press, Oxford,
  2004).

\bibitem{lorenz02}
T. Lorenz, M. Hofmann, M. Gruninger, A. Freimuth, G.~S. Uhrig, M. Dumm, and M.
  Dressel, Nature {\bf 418},  614  (2002).

\bibitem{auslaender05}
O.~M. Auslaender, H. Steinberg, A. Yacoby, Y. Tserkovnyak, B.~I. Halperin,
  K.~W. Baldwin, L.~N. Pfeiffer, and K.~W. West, Science {\bf 308},  88
  (2005).

\bibitem{blumenstein11}
C. Blumenstein, J. Sch\"afer, S. Mietke, A. Dollinger, M. Lochner, X.~Y. Cui,
  L. Patthey, R. Matzdorf, and R. Claessen, Nature Phys. {\bf 7},  776  (2011).

\bibitem{pham00}
K.-V. Pham, M. Gabay, and P. Lederer, Phys. Rev. B {\bf 61},  16397  (2000).

\bibitem{trauzettel04}
B. Trauzettel, I. Safi, F. Dolcini, and H. Grabert,
Phys. Rev. Lett. {\bf 92}, 226405 (2004).

\bibitem{lebedev05}
A.V. Lebedev, A. Cr\'epieux, and T. Martin,
Phys. Rev. B {\bf 71}, 075416 (2005).

\bibitem{pugnetti09}
S. Pugnetti, F. Dolcini, D. Bercioux, and H. Grabert,
Phys. Rev. B {\bf 79}, 035121 (2009).

\bibitem{das11}
S. Das and S. Rao,
Phys. Rev. Lett. {\bf 106}, 236403 (2011).

\bibitem{steinberg08}
H. Steinberg, G. Barak, A. Yacoby, L.~N. Pfeiffer, K.~W. West, B.~I. Halperin,
  and K.~L. Hur, Nature Phys. {\bf 4},  116  (2008).

\bibitem{imambekov09}
A. Imambekov and L.~I. Glazman, Science {\bf 323},  228  (2009).

\bibitem{imambekov09B}
A. Imambekov and L.~I. Glazman, Phys. Rev. Lett {\bf 102},  126405  (2009).

\bibitem{schmidt10}
T.~L. Schmidt, A. Imambekov, and L.~I. Glazman, Phys. Rev. Lett {\bf 104},
  116403  (2010).

\bibitem{shashi11}
A. Shashi, L.~I. Glazman, J.-S. Caux, and A. Imambekov, Phys. Rev. B {\bf 84},
  045408  (2011).
  
\bibitem{carmelo05}
J.~M.~P. Carmelo, K. Penc, and D. Bozi, Nucl. Phys. B {\bf 725},  421  (2005); 
{\bf 737},  351  (2006).

\bibitem{barak10}
G. Barak, H. Steinberg, L.~N. Pfeiffer, K.~W. West, L. Glazman, F. von Oppen,
  and A. Yacoby, Nature Phys. {\bf 6},  489  (2010).

\bibitem{white92}
S.~R. White, Phys. Rev. Lett {\bf 69},  2863  (1992).

\bibitem{white93}
S.~R. White, Phys. Rev. B {\bf 48},  10345  (1993).

\bibitem{white04}
S.~R. White and A.~E. Feiguin, Phys. Rev. Lett {\bf 93},  076401  (2004).

\bibitem{daley04}
A.~J. Daley, C. Kollath, U. Schollw\"ock, and G. Vidal, J. Stat. Mech.: Theor.
  Exp.  P04005  (2004).

\bibitem{schollwoeck05}
U. Schollw\"ock, Rev. Mod. Phys. {\bf 77},  259  (2005).

\bibitem{schollwoeck11}
U. Schollw\"ock, Ann. Phys. {\bf 326},  96  (2011).

\bibitem{bares90}
P.~A. Bares and G. Blatter, Phys. Rev. Lett {\bf 64},  2567  (1990).

\bibitem{bares91}
P.~A. Bares, G. Blatter, and M. Ogata, Phys. Rev. B {\bf 44},  130  (1991).

\bibitem{bares92}
P.~A. Bares, J.~M.~P. Carmelo, J. Ferrer, and P. Horsch, Phys. Rev. B {\bf 46},
   14624  (1992).

\bibitem{ogata91}
M. Ogata, M. Luchini, S. Sorella, and F. Assaad, Phys. Rev. Lett {\bf 66},
  2388  (1991).

\bibitem{moreno11}
A. Moreno, A. Muramatsu, and S.~R. Manmana, Phys. Rev. B {\bf 83},  205113
  (2011). 

\bibitem{soeffing12}
S. S\"{o}ffing, I. Schneider, and S. Eggert, http://arxiv.org/abs/1204.0003
  (2012).

\bibitem{carmelo06b}
J.~M.~P. Carmelo, L.~M. Martelo, and K. Penc, Nucl. Phys. B {\bf 737},  237
  (2006).

\bibitem{ogata90} M. Ogata and H. Shiba, 
                  Phys. Rev. B {\bf 41}, 2326 (1990).

\bibitem{lavalle03}
C. Lavalle, M. Arikawa, S. Capponi, F.~F. Assaad, and A. Muramatsu, Phys. Rev.
  Lett {\bf 90},  216401  (2003).

\end{thebibliography}
\end{document}